\documentclass[aps,prl,showpacs,amsmath,twocolumn,amssymb,superscriptaddress,letterpaper]{revtex4}
\usepackage{times}
\usepackage{amsfonts}
\usepackage{mathrsfs}
\usepackage{graphicx}
\usepackage{dcolumn}
\usepackage{bm}
\usepackage{color}

\usepackage[colorlinks,bookmarks=false,citecolor=blue,linkcolor=red,urlcolor=blue]{hyperref}

\usepackage{bibunits}
\begin{document}

\begin{bibunit}

\title{ Electronic structure properties in the nematic phases of FeSe }

\author{Yi Liang}
\affiliation{ Institute of Physics, Chinese Academy of Sciences,
Beijing 100190, China}

\author{Xianxin Wu}
\affiliation{ Institute of Physics, Chinese Academy of Sciences,
Beijing 100190, China}

\author{Jiangping Hu  }\email{jphu@iphy.ac.cn} \affiliation{ Institute of Physics, Chinese Academy of Sciences,
Beijing 100190, China}\affiliation{Department of Physics, Purdue University, West Lafayette, Indiana 47907, USA}
\affiliation{Collaborative Innovation Center of Quantum Matter, Beijing, China}

\date{\today}

\begin{abstract}
We investigate the electronic structures of FeSe in the presence of  different possible orders. We find that only ferro-orbital order (FO) and collinear antiferro-magnetism (C-AFM) can simultaneously induce splittings at $\Gamma$ and M. Bicollinear antiferro-magnetism (B-AFM) and spin-orbit coupling (SOC) have very similar band structure on $\Gamma$-M near the Fermi level. The temperature insensitive splitting at $\Gamma$ and the temperature dependent splitting at M observed in  recent experiments can be explained by the d-wave bond nematic (dBN) order together with SOC. The recent observed Dirac cones and their temperature ($T$) dependence in FeSe thin films can also be well explained by the dBN order with band renormalization. Their thickness- and cobalt-doping- dependent behaviors are the consequences of electron doping and reduction of Se height. All these suggest that the nematic order in FeSe system is the dBN order.
\end{abstract}


\maketitle

The  structurally simplest iron-based superconductor FeSe system has attracted enormous attention due to its many fantastic properties. The FeSe bulk without any doping shows superconductivity at $T_c=9$ K and $T_c$ can be significantly enhanced to 38 K by applying pressure \cite{CFelser2009}. Surprisingly, the $T_c$ can reach 65 K in 1UC FeSe on SrTiO$_3$ \cite{Wang2012,Tan2013,He2013,Zhang2014} and may be over 100 K \cite{JinFengJia2015}. Furthermore, an robust zero-energy bound state against magnetic field up to 8 T was observed at each interstitial iron impurity in superconducting Fe(Te,Se) and it bears all the characteristics of the Majorana bound state proposed for topological superconductors \cite{Pan2015}. In addition, nontrivial topological states have been predicted to exist in Fe(Te, Se) and 1UC FeSe films on SrTiO$_3$ substrates \cite{JiangpingHu2014,JiangpingHu2014-2, Wangzhijun2015}.

Recently, the nematic order in FeSe has attracted much attention. Nematicity, defined as the breaking of tetragonal rotational symmetry, is a well-established experimental fact in iron pnictides. Its origin is still highly debated between magnetic orders \cite{Fang2008,Qi2009,Cano2010,Liang2013} and orbital orders \cite{Lee2009,Onari2012,Yamase2013,Stanev2013}. The former is strongly supported by the facts that the orthorhombic lattice distortion is always accompanied by the collinear magnetic order in iron pnictides. However, the bulk FeSe shows an orthorhombic lattice distortion at $T_s\sim90$ K without any evidence of magnetic phase transition. A band splitting at M is observed in FeSe below $T_{nem}\sim120$ K by angle-resolved photoemission spectroscopy (ARPES) \cite{HDing2014,HDing2015,Coldea201501,Coldea201502,DHLu2015-1,DLFeng2015-1,ZXShen2015}, which is taken as the direct signature of the nematicity. Very recently, the T-insensitive splitting at $\Gamma$ and T-sensitive splitting at M have been reported \cite{HDing2015}. What is more, Dirac cones have been discovered around M point in nematic phase of FeSe thin films and cobalt doping can suppress the nematicity \cite{DLFeng2015-1,ZXShen2015}. So far, no strong evidence of anisotropy has been obtained in FeSe as $T_s<T<T_{nem}$. As the splitting at high symmetry points are highly confined by symmetry, it is possible for us to seek the constraints on different orders in order to consistently explain all experimental observations in the FeSe nematic phase.

In this paper, we try to answer the above questions and understand the nematicity in FeSe system, which not only can help us to understand the origin of the fantastic properties of FeSe, but also may shade much light on the  superconducting mechanism of iron-based superconductors (FeSCs). We, with the five-orbital tight-binding (TB) model for single unit-cell (UC) FeSe films, investigate the effects on the band structure of eight possible orders in FeSe: FO order, antiferro-orbital (AFO) order, dBN order, charge order (CO), C-AFM, B-AFM, N\'{e}el antiferro-magnetism (N-AFM) and SOC. We find that only FO and C-AFM can simultaneously induce the splittings at $\Gamma$ and M, but  in the C-AFM state, it is accompanied by additional band folding. The splitting at $\Gamma$ can be induced in B-AFM and SOC. These two orders, near Fermi energy ($E_{F}$), have very similar band structure on $\Gamma$-M, but in B-AFM phase, two electron pockets must emerge at $X$. The splitting at M can arise in the presence of the dBN order. The left other orders can not produce the splittings observed in ARPES. The different $T-$dependence of the splitting at $\Gamma$ and M can be explained by the dBN order together with SOC. The   recently observed Dirac cones and their T-dependence in FeSe thin films can also be well explained by the dBN order with band renormalization. The cobalt-doping- and thickness- dependent behaviors result from the electron doping, as well as the reduction of Se height. All these suggest that the nematic order in FeSe system is dominated by the dBN order.

\begin{figure}[t]
\centerline{\includegraphics[width=8 cm]{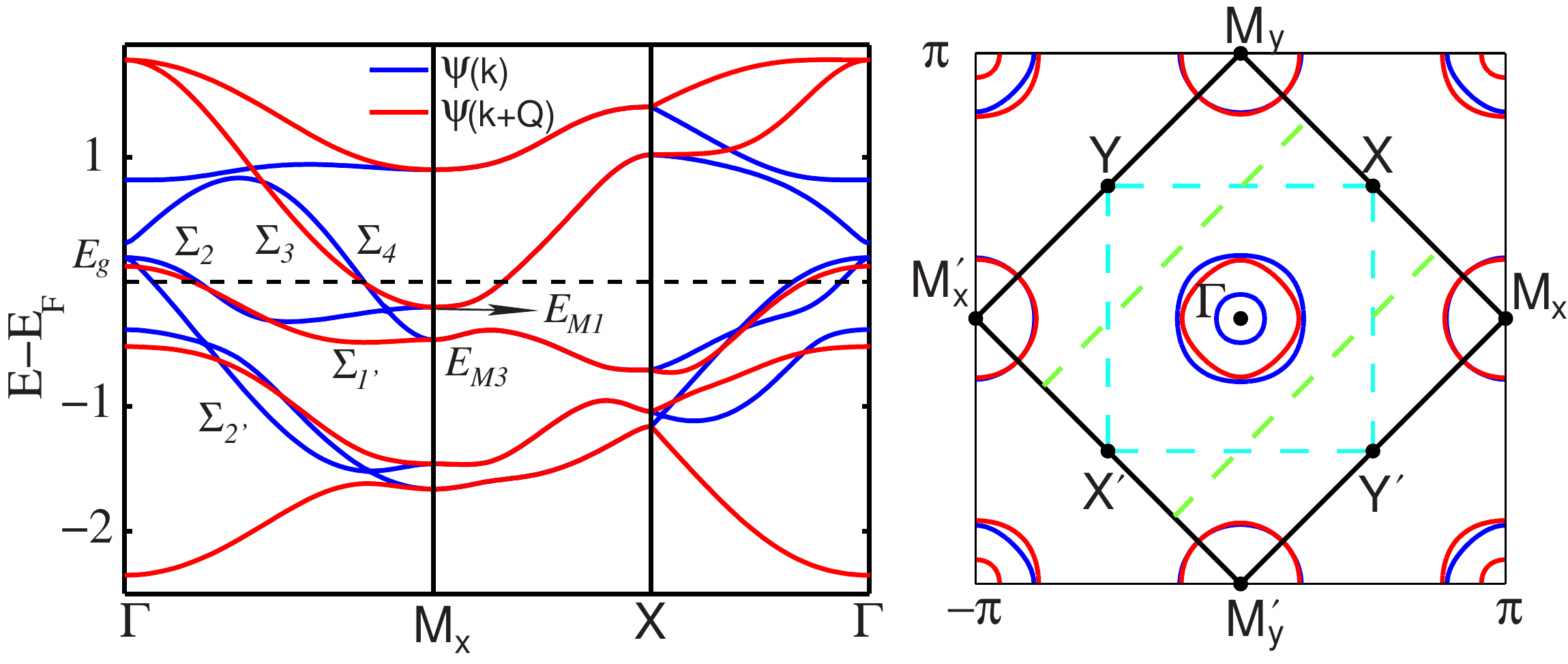}}
\caption{Band structure of 1UC FeSe film without any ordering (a) and its corresponding Fermi surface in the Brillouin zone (BZ) for Fe lattice (b). The bands attributed to $\psi(\mathbf{k})$ and $\psi(\mathbf{k+Q})$ is plotted in red and blue respectively. The cyan dotted square and the green dashed rectangle are the BZ for C-AFM and B-AFM orders respectively.}
\label{band_BZ}
\end{figure}

{\it Effects of orders on electronic structure: }
To investigate the effects of different orders, we start from the five-orbital TB model for 1UC FeSe films with the lattice parameters of bulk FeSe. The results have no qualitative difference with those from the TB model for bulk. The Hamiltonian containing five Fe $3d$ orbitals is given by,

\begin{eqnarray}
H_{0}=\sum_{\mathbf{k}\in BZ1,\sigma}\psi_{\sigma}^{\dag}(\mathbf{k})A_0(\mathbf{k})\psi_{\sigma}(\mathbf{k}),
\end{eqnarray}
where $\psi^{\dag}_{\sigma}(\mathbf{k}) =[C_{\mathbf{k}1\sigma}^{\dag},C_{\mathbf{k}2\sigma}^{\dag},C_{\mathbf{k+Q}3\sigma}^{\dag},C_{\mathbf{k+Q}4\sigma}^{\dag},C_{\mathbf{k+Q}5\sigma}^{\dag}]$
, $\mathbf{Q}=(\pi,\pi)$, $A_0(\mathbf{k})$ is given in supplement materials and BZ1 denotes the Brillouin zone of one Fe lattice. Here, the natural gauge is taken and the Fe $3d$ orbitals, for convenience, are designated as numbers, {\em i.e.}, $d_{xz}\rightarrow1,d_{yz}\rightarrow2,d_{x^2-y^2}\rightarrow3,d_{xy}\rightarrow4,d_{z^2}\rightarrow5$. The band structure of the above model is shown in Fig. \ref{band_BZ}(a). The bands from $\psi(\mathbf{k})$ are plotted in blue and those from $\psi(\mathbf{k+Q})$ in red.  The degeneracy at $\Gamma$ results from the equivalence of $d_{xz}$ and $d_{yz}$ orbitals and no coupling between them at $\Gamma$, which is protected by $S_4$ symmetry. Thus, there are two ways to break the degeneracy : one is to break the $S_4$ symmetry on Fe sites as well as the $C_4$ symmetry on Se sites; the other is to induce coupling between $d_{xz}$ and $d_{yz}$ orbital at $\Gamma$.
As for the two-fold degenerate states (excluding spin degeneracy) on M-X, one is the $\psi(\mathbf{k})$ band and the other is $\psi(\mathbf{k+Q})$ band. From the point of view of symmetry, the band degeneracy on M-X is protected by the symmetry of $\Upsilon=K\{C_{2Y}|T_{x/y}\}$, where $K$ is the conjugate operator and $C_{2Y}$ is the $C_{2}$ rotation operator along the diagonal of Fe lattice and $T_{x/y}$ is the translation in the $x/y$ direction by the Fe-Fe distance. The anti-unitary operator $\Upsilon$ commutes with $H_0$ and $\Upsilon^2=-1$ on $\Upsilon$-invariant M-X line, which infers that the bands on M-X are two-fold degenerate.  Hence, the effective way to remove the degeneracy on M-X is to break $\Upsilon$ symmetry. Note that small band splitting on M-X may alter the superconducting pairing symmetry due to the change of the  topology of electron Fermi pockets from two intersecting ellipses into two separated concentric electron pockets.
On $\Gamma$-M$_{x/y}$ line,  $d_{xz/yz}$ and $d_{xy}$ orbitals only couple with each other and the other three orbitals hybridize. The three bands across $E_{F}$ on $\Gamma$-M$_x$, $\Sigma_{2},\Sigma_{2'}$ and $\Sigma_{1'}$, consist of $\{C_{\mathbf{k},2},C_{\mathbf{k+Q},3},C_{\mathbf{k+Q},5}\},$$\{C_{\mathbf{k},1},C_{\mathbf{k+Q},4}\}$
and $\{C_{\mathbf{k},4},C_{\mathbf{k+Q},1}\}$ respectively. The band hybridization among these three bands can be induced by any coupling between their components. Near the M point, the two bands from $\psi(\mathbf{k})$, $\Sigma_{2}$ and $\Sigma_{4}$, form a Dirac cone. In the following, we discuss the effects of different orders and SOC on the band structure.

\begin{figure}[t]
\centerline{\includegraphics[width=8 cm]{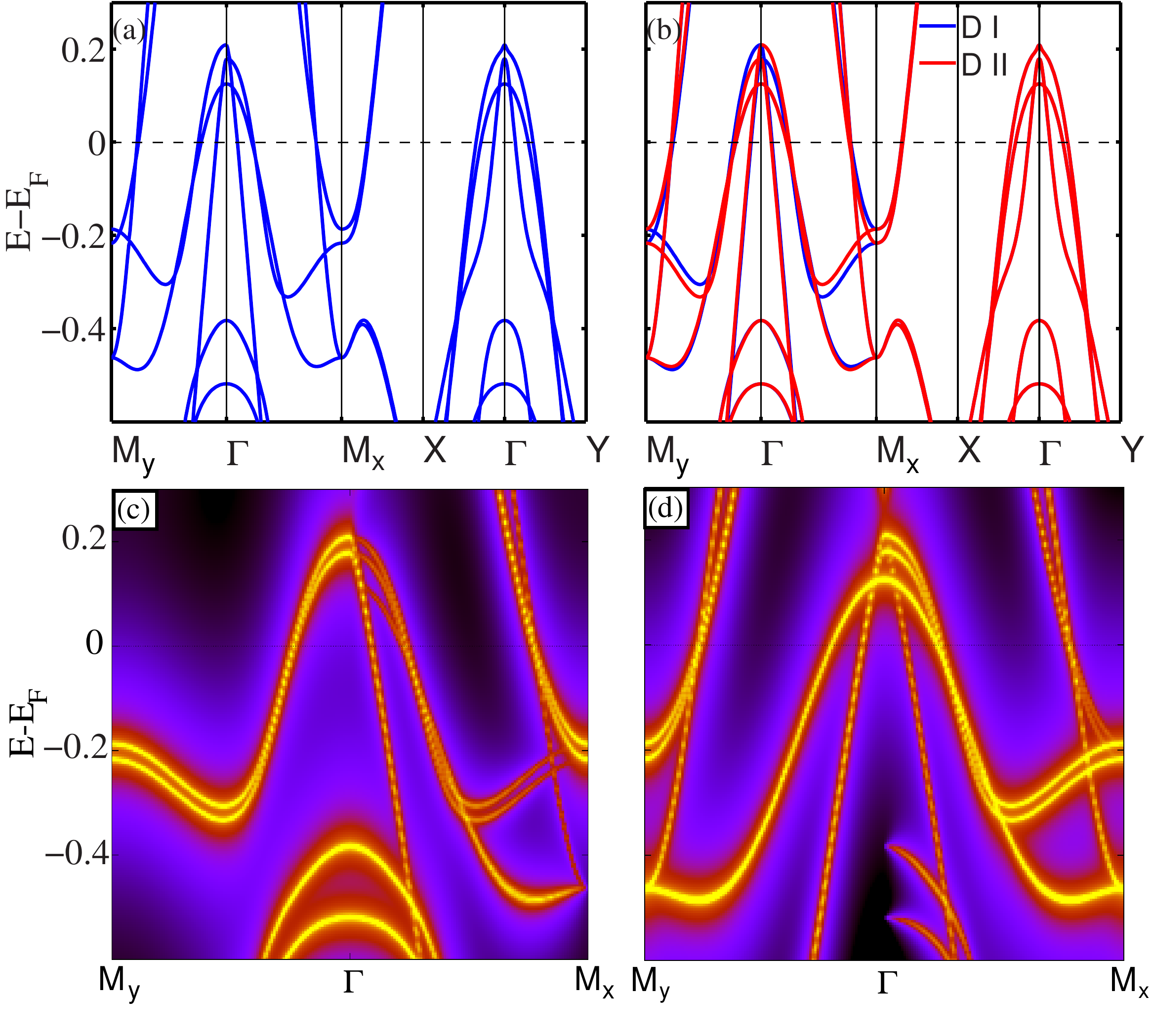}}
\caption{Band structure with FO order (a) and that for a twinned sample (b), and the polarization-dependent bands in ARPES with respect to $\Gamma$-M$_x$ direction: (c) for even orbitals and (d) for odd orbitals. $\Delta_{FO}=30$ meV, to induce the observed splitting at $\Gamma$ \cite{HDing2015}. In (b), the bands attributed to domain D I and domain D II  are plotted in red and blue respectively. }
\label{bandFO}
\end{figure}

\textit{A. Ferro-orbital order: } FO order, which is induced by tetragonal symmetry breaking, is characterized by the orbital polarization between $d_{xz}$ and $d_{yz}$ orbitals. The additional Hamiltonian term induced by this order can be written as,
\begin{equation}
h^{FO}=\Delta_{FO}\sum_{k\in BZ1,\sigma}(n_{k,1,\sigma}-n_{k,2,\sigma}),
\end{equation}
where $n_{k,\alpha,\sigma}=C_{k,\alpha,\sigma}^{+}C_{k,\alpha,\sigma}$.
Fig. \ref{bandFO}(a) shows the band structure with FO order. The most striking feature is the $2\Delta_{FO}$ gaps opened simultaneously at $\Gamma$ and M points between the $d_{xz}$ and $d_{yz}$ bands.
The band splitting on M$_{x/y}$-X results from the symmetry breaking of $\Upsilon$.
Fig. \ref{bandFO}(b) also provides the band structure of a twinned sample composed of two orthogonal domains of which the order parameters are opposite, {\em i.e.}, $\Delta_{FO}^{DII}=-\Delta_{FO}^{DI}$. Since generally, the beam spot size of incident light in ARPES is larger than the domain size, the band structure observed in ARPES is the combination of the bands for the two domains. As the two domains are connected by a $C_4$ rotation, the splitted states become degenerate again and the bands on M$_{y}$-$\Gamma$-M$_{x}$ reappear symmetric. However, the symmetry breaking can also been clearly observed in the polarization dependent ARPES measurements, as shown in Fig. \ref{bandFO}(c) for the even orbitals and Fig. \ref{bandFO} (d) for the odd orbitals.

\textit{B. Antiferro-orbital order:} AFO order is also characterized by the orbital polarization between $d_{xz}$ and $d_{yz}$ orbitals but the polarization changes alternatively with Fe sites.
It can be defined as $\epsilon_{j,xz}-\epsilon_{j,yz}=\Delta_{AFO}e^{i\mathbf{Q}\cdot\mathbf{R}_{j}}$ where  $\epsilon_{j,\alpha}$ is the on-site energy of orbital $\alpha$ on site $j$ and $\mathbf{R}_{j}$ is the position of site $j$. Thus the additional AFO Hamiltonian term, in momentum space, can be written as
\begin{equation}
h^{AFO}=\Delta_{AFO}\sum_{k\in BZ1,\sigma,\alpha=1,2}
   sgn(\alpha) C_{\mathbf{k},\alpha,\sigma}^{+}C_{\mathbf{k+Q},\alpha,\sigma}
\end{equation}
where $sgn(\alpha)$ equals 1 for $\alpha=1$ and -1 for $\alpha=2$.

AFO order breaks the $S_{4}$ symmetry on Fe site and inversion symmetry, but preserves the $C_{4}$ symmetry on Se sites and  translational symmetry. Thus, although this order induces the coupling between $\psi(\mathbf{k})$ and $\psi(\mathbf{k+Q})$ bands, there is no degeneracy removal at $\Gamma$ and M points except energy shifts for $d_{xz/yz}$ bands. The most distinct feature with AFO is the hybridization between $\Sigma_{1'}$ and $\Sigma_{2'}$ but no hybridization  between $\Sigma_{1'}$ and $\Sigma_{2}$.

\textit{C. $d$-wave bond nematic order:} The dBN order is characterized by the hopping difference between $x$ and $y$ direction for $d_{xz/yz}$ orbitals \cite{HDing2015, TLi2015, Andersen2015}. It can originate from lattice orthorhombic distortion or interatomic Coulomb repulsion \cite{ZiQiangWang2015}. The additional Hamiltonian introduced by dBN order is given by,

\begin{eqnarray}
h^{dBN} &=& \sum_{i,\sigma,\alpha=1,2}\frac{\Delta_{dBN}}{8}[C_{i,\alpha,\sigma}^{+}C_{i\pm e_{x},\alpha,\sigma}-C_{i\alpha\sigma}^{+}C_{i\pm e_{y}\alpha\sigma}]\nonumber\\
 &=& \sum_{\mathbf{k}\in BZ1,\sigma,\alpha=1,2}\frac{\Delta_{dBN}}{4}(cosk_{x}-cosk_{y})n_{k,\alpha,\sigma}.
\end{eqnarray}

\begin{figure}[t]
\centerline{\includegraphics[width=8 cm]{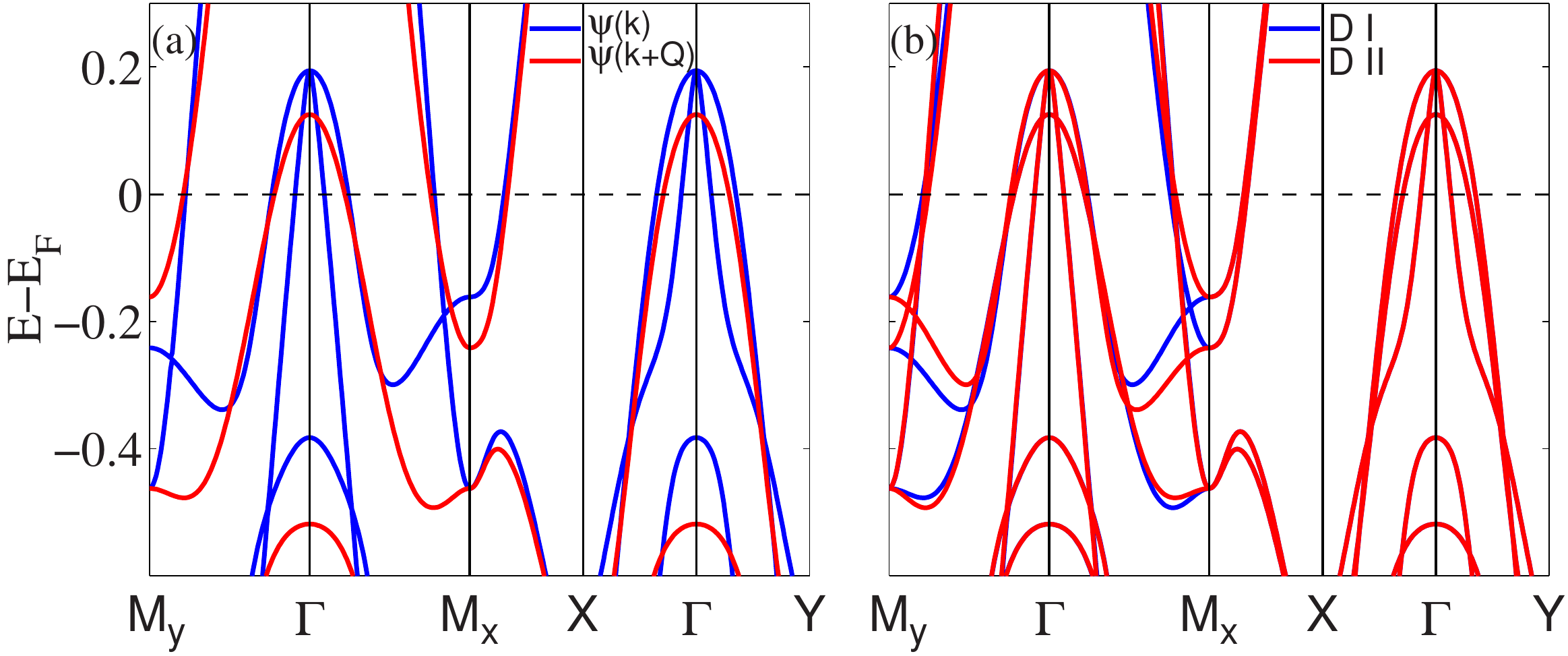}}
\caption{Band structure with dBN order (a) and that for a twinned sample (b). $|\Delta_{dBN}|=80$ meV, the splitting observed at M in FeSe thin films \cite{DHLu2015-1}.}
\label{banddBN}
\end{figure}

In dBN order, the $S_4,C_4$ and $\Upsilon$ symmetries are broken but the glide symmetry is preserved. Therefore, the band degeneracy on M-X is removed and $\psi(\mathbf{k})$ and $\psi(\mathbf{k+Q})$ bands are still decoupled. As the dBN term vanishes on $\Gamma-X/Y$ and achieves the maximum at M$_{x/y}$, an splitting of $\Delta_{dBN}$ is induced at M$_{x/y}$ but on splitting at $\Gamma$ and $X/Y$ points. Specifically, in dBN order, the $\psi(\mathbf{k})$ and $\psi(\mathbf{k+Q})$ bands attributed to $d_{xz/yz}$ orbitals have  energy shifts of $-\frac{\Delta_{dBN}}{2}$ and $\frac{\Delta_{dBN}}{2}$ at $M_x$ respectively. These bands at $M_y$ show the opposite shift behavior, which can be seen in Fig. \ref{banddBN}(a). Considering the possible domains in experiments, we also provide the band structure of a twinned system with dBN order in Fig. \ref{banddBN}(b).

\begin{figure}[t]
\centerline{\includegraphics[width=8 cm]{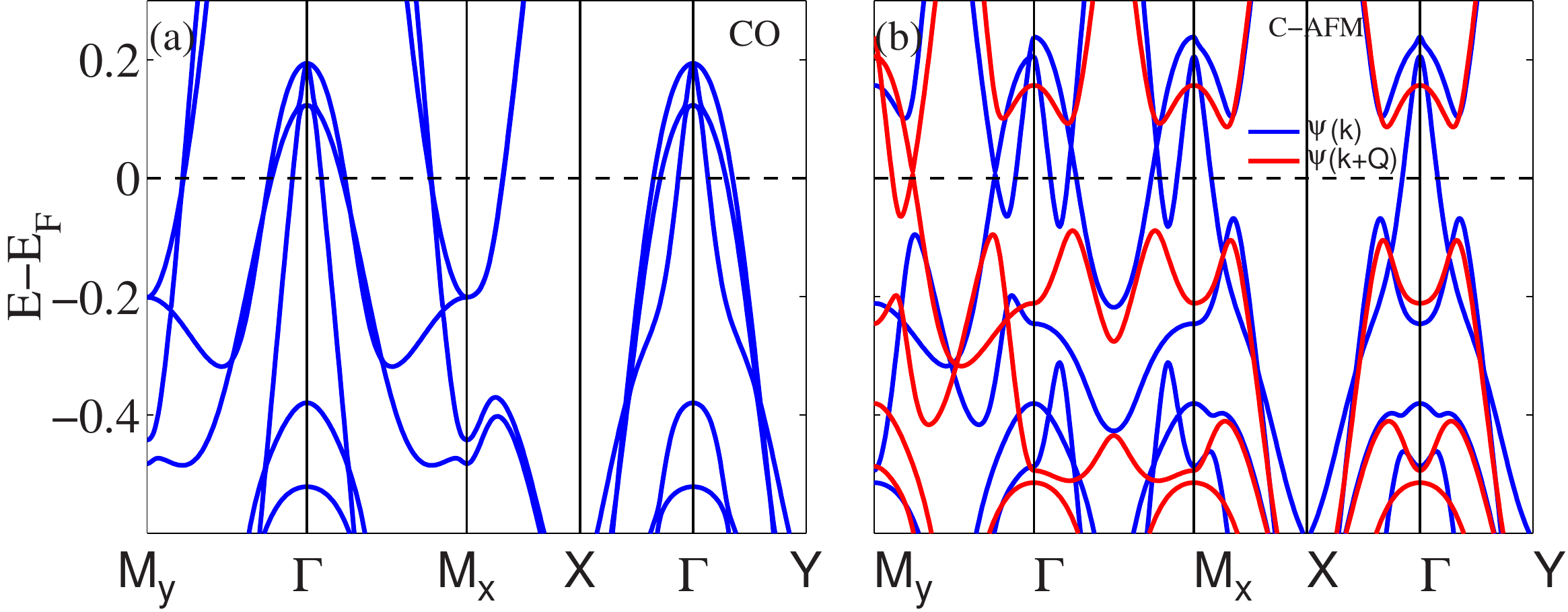}}
\caption{Band structure with CO order (a) and that with C-AFM order (b). $\Delta_{CO}=40$ meV. To induce a splitting of $30$ meV at $\Gamma$,  $\Delta_{CAFM}=140$ meV.}
\label{CO_CAFM}
\end{figure}

\textit{D. Charge order: }
Earlier theoretical study suggested that, in FeSCs, spin-density waves (SDW)
can induce charge-density waves (CDW) with the modulation momentum, $\mathbf{q}_{CDW}$,
double of wave vector of SDW, $\mathbf{q}_{SDW}$ \cite{JianXinZhu2010}. Thus,
$\mathbf{q}_{CDW}=(\pi,\pi)$ in FeTe. In FeAs system, besides $\mathbf{q}_{CDW}=(0,0)$,
a CDW with $\mathbf{q}_{CDW}=(\pi,\pm\pi)$ can also be caused at boundaries of SDW domains. Although no long range magnetic order is observed in FeSe, it's believed that it's the result of competition of different magnetic fluctuations, which has been demonstrated by neutron scattering and nuclear magnetic resonance \cite{BBuchner2015,CMeingast2015,ATBoothriyd2015,JZhao2015}. Thus, we can assume that a CDW with $\mathbf{q}_{CDW}=(\pi,\pi)$ exists in the normal state of FeSe.
The $(\pi,\pi)$ CDW is introduced by the difference of on-site energy
on two sublattice of Fe, $\epsilon_{\alpha}^{A}-\epsilon_{\alpha}^{B}=2\Delta_{CDW}$.
The additional Hamiltonian term is
\begin{equation}
  h^{CDW}=\Delta_{CDW}\sum_{\mathbf{k}\in BZ_{1}}C_{\mathbf{k},\alpha,\sigma}^{+}C_{\mathbf{k}+\mathbf{Q},\alpha,\sigma}.
\end{equation}
The $C_{4}$ and $\Upsilon$ symmetries are broken, but $S_{4}$ symmetry is preserved. Therefore, no anisotropy occurs in $x,y$ direction and the band degeneracy on M-X is removed, as shown in Fig. \ref{CO_CAFM}(a). The most distinct features are the splitting of  $E_{M3}$ states (labeled in Fig.1), which results from the coupling between $C_{k,4,\sigma}$ and $C_{k+Q,4,\sigma}$. In addition, bands $\Sigma_{1'}$and $\Sigma_{2'}$ hybridize near their intersection point.

\textit{E. Collinear AFM:}
AFM fluctuations have been found by neutron scattering and nuclear magnetic resonance in FeSe \cite{BBuchner2015,CMeingast2015,ATBoothriyd2015,JZhao2015}. A DFT calculation also suggested close competition between C-AFM and B-AFM in FeSe \cite{Ma2009}. Thus in the following, we consider the effects on band structure of three magnetic orders: C-AFM, B-AFM and N\'{e}el AFM. Firstly, let's discuss the effects of the most popular AFM order in FeSCs, C-AFM. It can be introduced by a spin polarized term, $\epsilon_{j,\alpha,\uparrow}-\epsilon_{j,\alpha,\downarrow}=2e^{i\mathbf{q}_{CA}\cdot\mathbf{R}_{j}}\Delta_{CA}$($\mathbf{q}_{CA}=(\pi,0)$),
which leads to an additional Hamiltonian term,
\begin{equation}
h^{CA} = \Delta_{CA}\sum_{\mathbf{k}\in BZ1,\alpha,\sigma}
  sgn(\sigma)C_{\mathbf{k},\alpha,\sigma}^{+}C_{\mathbf{k}+\mathbf{q}_{CA},\alpha,\sigma}.
\end{equation}
where $sgn(\sigma)$ is the sign function, equals 1 for spin up and -1 for spin down.
Fig. \ref{CO_CAFM}(b) presents the band structure with C-AFM which is much different from that of the normal state. C-AFM doubles the unit cell and rotates it by 45 degrees, which reduces the volume of BZ for 2Fe unit cell by one half and induces band folding and Fermi surface reconstruction.
The normal-state bands located out of the magnetic BZ  are folded into the magnetic BZ with respect to its boundary. The normal-state bands around M are folded onto $\Gamma$ and then new bands near $E_F$ appears around $\Gamma$. M point turns equivalent to $\Gamma$ in C-AFM. The splitting at $\Gamma$ point $\Delta_{E_{g}}\propto \Delta_{CA}^2$ from perturbation theory \cite{note01}.

\begin{figure}[t]
\centerline{\includegraphics[width=8 cm]{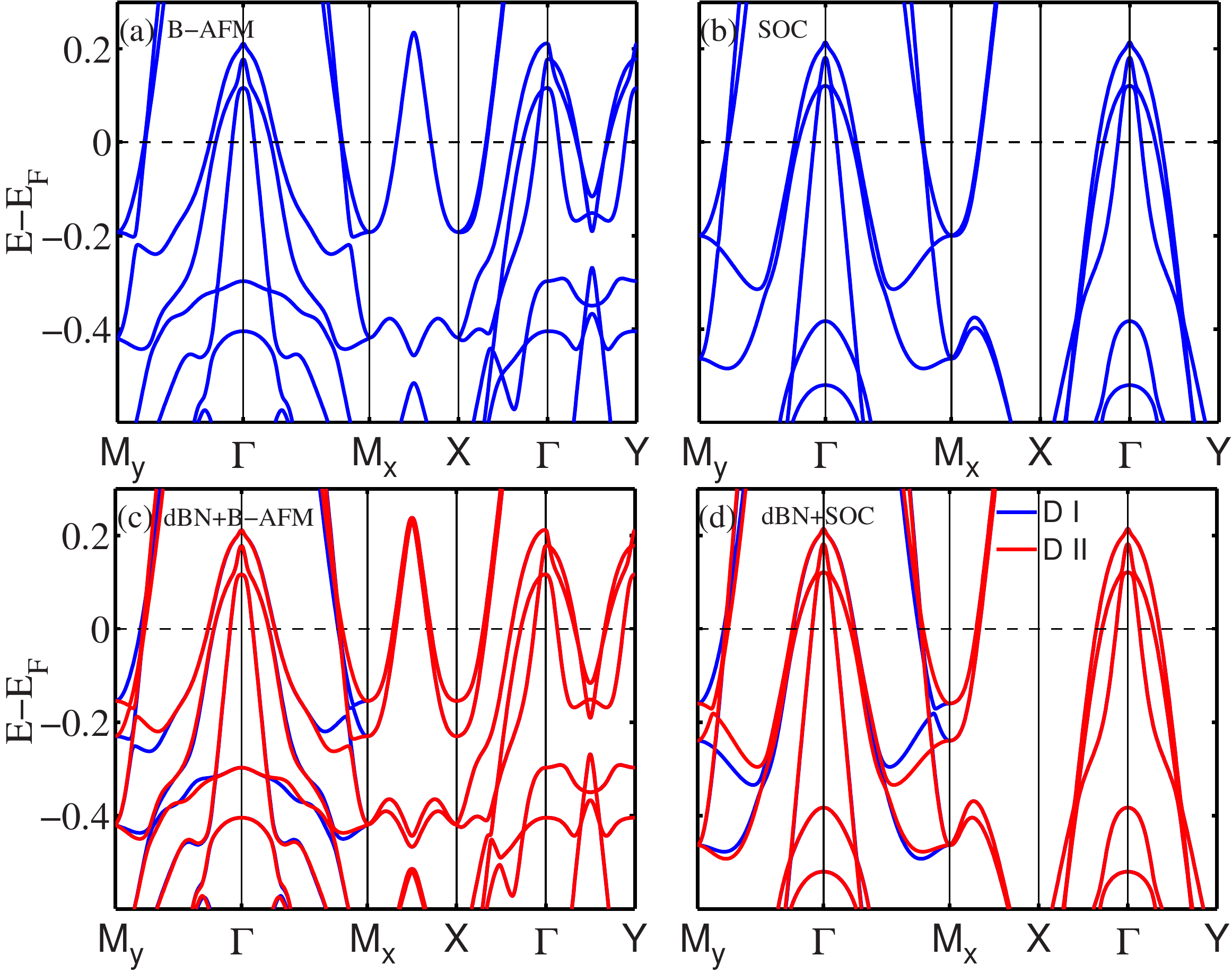}}
\caption{Band structures for a sample with SOC (a)and B-AFM order (b), and those for a twinned sample with coexistence of SOC and dBN order (c) and coexistence of B-AFM and dBN orders (d). In the four cases, $\lambda=30$ meV, $\Delta_{BA}=220$ meV, $\Delta_{dBN}=80$ meV, such that the splitting at $\Gamma$ and M are about $30$ and 80 meV respectively, the values observed in ARPES \cite{HDing2015,DHLu2015-1}. }
\label{SOC_BAFM_dBN}
\end{figure}

\textit{F. Bicollinear AFM:}
Now we consider the effects of B-AFM on band structure and it can be modeled by $\epsilon_{j,\alpha,\uparrow}-\epsilon_{j,\alpha,\downarrow}=
  e^{i\mathbf{q}_{BA}\cdot\mathbf{R}_{j}}\Delta_{BA}[1+e^{i\mathbf{Q}\cdot\mathbf{R}_{j}}-i(1-e^{i\mathbf{Q}\cdot\mathbf{R}_{j}})]$
where $\mathbf{q}_{BA}=(\frac{\pi}{2},-\frac{\pi}{2})$. The additional Hamiltonian term is

\begin{eqnarray}
h^{BA} &=& \Delta_{BA}\sum_{k\in BZ1,\alpha,\sigma}
(\frac{1-i}{2}C_{\mathbf{k},\alpha,\sigma}^{+}C_{\mathbf{k}+\mathbf{q}_{BA},\alpha,\sigma}\nonumber\\
   & &
   +\frac{1+i}{2}C_{\mathbf{k},\alpha,\sigma}^{+}C_{\mathbf{k}-\mathbf{q}_{BA},\alpha,\sigma})sgn(\sigma).
\end{eqnarray}
In B-AFM, the $S_4,C_4$ and translational symmetries are broken but $\Upsilon$ symmetry survives. Thus, the $d_{xz/yz}$ bands split at $\Gamma$ point and bands on M$_{x/y}$-X are still degenerate, as shown in Fig. \ref{SOC_BAFM_dBN}(a). The splitting $\Delta_{E_g}=0.36\Delta_{BAFM}^{2}$ in our model \cite{note01}. In addition, Band $\Sigma_{1'}$ hybridizes with $\Sigma_{2},\Sigma_{2'}$ due to the indirect hopping between $C_{\mathbf{k},\alpha,\sigma}^{+}$ and  $C_{\mathbf{k+Q},\alpha,\sigma}$ through $C_{\mathbf{k\pm q}_{BA},\alpha,\sigma}$. Another striking effect is the emergence of two additional electron pockets at $X$, which is induced by band folding. The magnetic primitive cell is a $2\times1$ (or $1\times2$) supercell of 2Fe unit cell. The magnetic BZ becomes a rectangle and the rectangle in our case is shown with green dash line in Fig. \ref{band_BZ}(b). The normal-state bands out of magnetic BZ  are folded into it with respect to its boundary. In our cases, the bands around $Y-M_{y}$ are folded onto $\Gamma-X$, which gives rise to the two additional electron pockets at $X$ and the symmetric band structures on M-X and $\Gamma-Y$ with respect to their centers.

\textit{G. N\'{e}el AFM: }
As for the N-AFM order, it can be modeled by, $\epsilon_{i,\alpha,\uparrow}-\epsilon_{i,\alpha,\downarrow}=2\Delta_{NA}e^{i\mathbf{Q}\cdot\mathbf{R}_{i}}$.
The additional term  introduced by the N-AFM order is,
\begin{equation}
h^{NA}=\Delta_{NA}\sum_{\mathbf{k}\in BZ1,\alpha}
  (C_{\mathbf{k}\alpha\uparrow}^{+}C_{\mathbf{k+Q}\alpha\uparrow}
   -C_{\mathbf{k}\alpha\downarrow}^{+}C_{\mathbf{k+Q}\alpha\downarrow}).
\end{equation}
In N-AFM phase, the magnetic unit cell is the same as that in the normal state, so no band folding arises. The band structure is very similar with that of CO. The $d_{xy}$ bands splits at M, the band degeneracy on M-X is removed and bands $\Sigma_{1'}$ and $\Sigma_{2'}$ hybridize near their intersection point.

\textit{H. SOC effects: }
Finally, we analyze the SOC effects on electronic structure of FeSe. Due to the inversion symmetry with respect to bond centers of nearest Fe-Fe, we consider an isotropic on-site SOC, $h^{SOC}=\lambda\sum_{i}\mathbf{L}_{i}\mathbf{S}_{i}$ where $\sum_{i}$ sums over the Fe sites. In supplement materials can find the detailed expression of $h^{SOC}$ in $3d-$orbital space. Fig. \ref{SOC_BAFM_dBN}(b) illustrates the band structure with SOC. On-site SOC doesn't break time reversal and space symmetries, only leads to $\Upsilon$ symmetry breaking and hybridization of orbitals over the whole $BZ1$. Thus, the band structure still has $C_4$ rotational and inversion symmetries and the degenerate bands on M-X are splitted into two branches. Band $\Sigma_{1'}$ hybridizes with bands $\Sigma_{2},\Sigma_{2'}$ due to the couplings of $C_{\mathbf{k},1/2,\sigma}$ and $C_{\mathbf{k},3/4/5,\bar{\sigma}}$. The splitting at $\Gamma$ is the result of the coupling of $d_{xz}$ and $_{yz}$ at $\Gamma$. The splitting $\Delta_{E_g}=\lambda+a\cdot \lambda^2$ where $a$ is an model-dependent parameter \cite{note01}.

\begin{table}
\caption{Effects of different orders. $\Delta$ denotes a splitting with the
order of $\Delta_{\alpha}$ and $\Delta^{2}$ for that with the order of $\Delta^{2}_{\alpha}$, where $\Delta_{\alpha}$ is the order parameter of order $\alpha$. $\checkmark$ means band spltting and $\mathsf{X}$ means no band splitting. }
\begin{tabular}{ccccccc}
\hline
 & \multicolumn{4}{c|}{splitting} & \multicolumn{2}{c}{hybridization}\tabularnewline
\hline
 & $E_{g}$ & $E_{M1}$ & $E_{M3}$ & \multicolumn{1}{c|}{M-X} & $\Sigma_{1'}$and $\Sigma_{2}$  & $\Sigma_{1'}$and $\Sigma_{2'}$\tabularnewline
\hline

SOC & $\Delta$ & $\mathsf{X}$ & $\mathsf{X}$ & $\checkmark$ & $\checkmark$ & $\checkmark$\tabularnewline

FO & $\Delta$ & $\Delta$ & $\mathsf{X}$ & $\checkmark$ & $\mathsf{X}$ & $\mathsf{X}$\tabularnewline

AFO & $\mathsf{X}$ & $\mathsf{X}$ & $\mathsf{X}$ & $\mathsf{X}$ & $\mathsf{X}$ & $\checkmark$\tabularnewline

dBN & $\mathsf{X}$ & $\Delta$ & $\mathsf{X}$ & $\checkmark$ & $\mathsf{X}$ & $\mathsf{X}$\tabularnewline

CO & $\mathsf{X}$ & $\mathsf{X}$ & $\Delta$ & $\checkmark$ & $\mathsf{X}$ & $\checkmark$\tabularnewline

C-AFM & $\Delta^{2}$ & $\Delta^{2}$ & $\mathsf{X}$ & $\checkmark$ & $\mathsf{X}$ & $\mathsf{X}$\tabularnewline

B-AFM & $\Delta^{2}$ & $\mathsf{X}$ & $\mathsf{X}$ & $\mathsf{X}$ & $\checkmark$ & $\checkmark$\tabularnewline

N-AFM & $\mathsf{X}$ & $\mathsf{X}$ & $\Delta$ & $\checkmark$ & $\mathsf{X}$ & $\checkmark$\tabularnewline
\hline
\end{tabular}
\label{Tab_Effects}
\end{table}

\textit{Discussion: }
The effects of eight considered orders on band structure have been summarized in Table \ref{Tab_Effects}. Four orders can induce the splitting of $E_{g}$ states and three ones can produce the splitting of $E_{M1}$ states. Only FO and C-AFM orders simultaneously break the degeneracies of $E_{g}$ and $E_{M1}$ states. C-AFM causes the normal-state bands around $\Gamma$ and M$_{x/y}$ to fold onto each other. SOC and B-AFM have very similar effects, as shown in Table \ref{Tab_Effects}. However, B-AFM induces the normal-state bands around $\Gamma$-X and Y-M$_{y}$ to fold onto each other and then two additional electron  pockets appear at X. Furthermore, a much larger order parameter is needed in B-AFM to produce the same splitting at $\Gamma$ compared with that in SOC, since the splitting is an second order effect in B-AFM. It's difficult to distinguish CO and N-AFM from the band structure, the measurement of charge distribution or magnetic moment on each Fe site is needed. In all the considered orders, SOC and B-AFM order can simultaneously generate hybridization between $\Sigma_{1'}$ and $\Sigma_{2/2'}$. No order can produce hybridization between $\Sigma_{1'}$ and $\Sigma_{2}$ but no hybridization between $\Sigma_{1'}$ and $\Sigma_{2'}$. While dBN coexists with SOC or B-AFM, the band structures for twinned samples  are very similar with that for a twinned sample with FO order, at least on $\Gamma$-M, as shown in Fig. \ref{SOC_BAFM_dBN}(c,d). The difference is that, in the complex cases, the splittings at $\Gamma$ and M can be different and only the splitting at M is $T$-dependent while in FO, both are $T$-dependent and have the same values. The complex cases of dBN+SOC or dBN+B-AFM can almost explain the the ARPES observation by Zhang {\em et al.} \cite{HDing2015} except the little hybridization between $\Sigma_{1'}$ and $\Sigma_{2'}$. Considering the observed Dirac cones, the dBN+B-AFM case is  excluded, because the dramatic breaking of Dirac cones by B-AFM, as shown in Fig. \ref{SOC_BAFM_dBN}(c). This assertion is based on our following explanation of the origin of the observed Dirac cones in ARPES.

\begin{figure}[t]
\centerline{\includegraphics[width=8 cm]{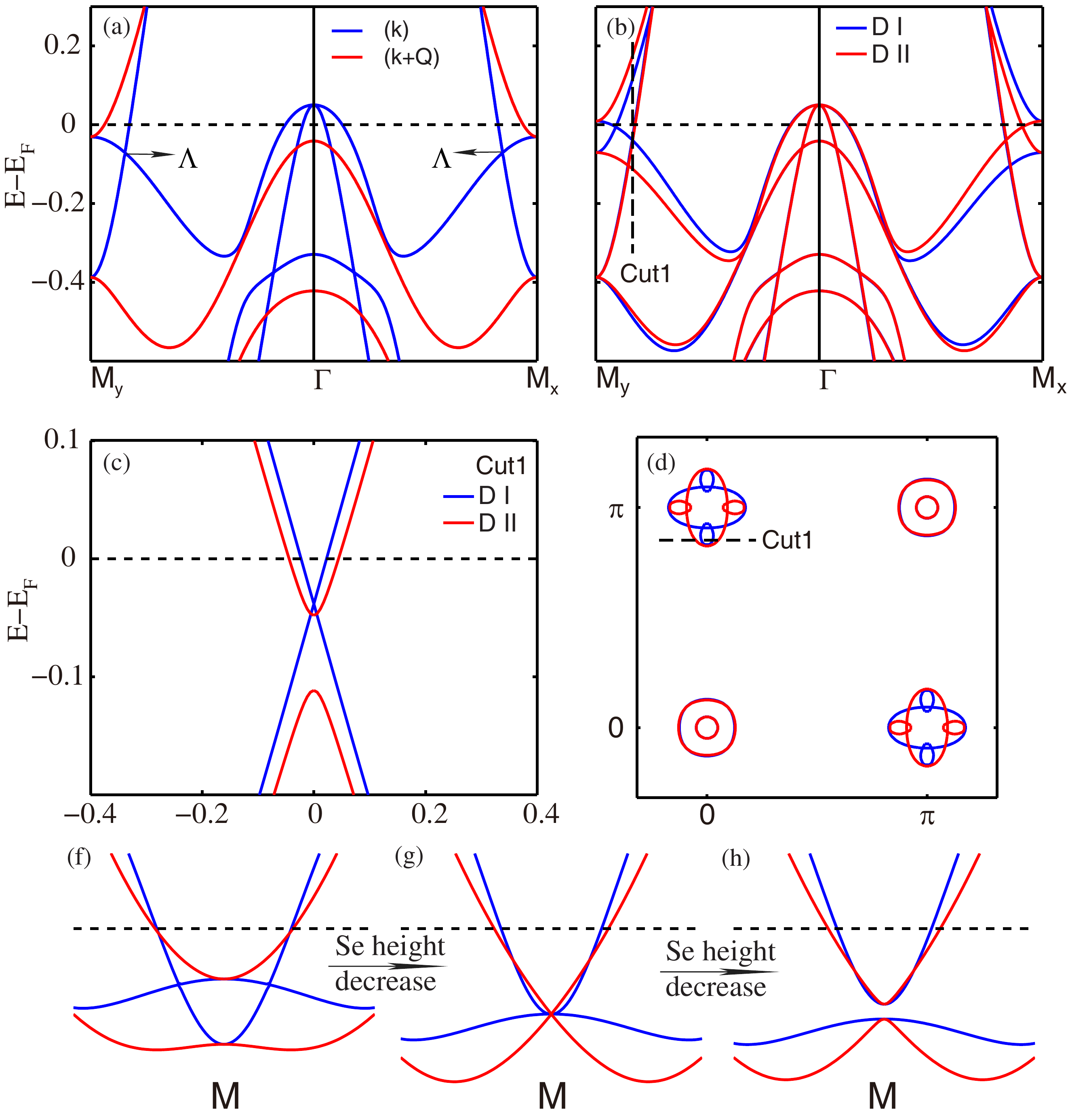}}
\caption{(a) Band structure renormalized by nearest-neighbor Coulomb interaction $H_V$ with $V=0.55eV$, $H_{V}=V\sum_{\langle i,j\rangle}n_i n_j$ where $n_i=\sum_{\alpha}n_{i,\alpha}$. $\Lambda$ is a Dirac point. (b) Band structure with dBN order for a twinned sample. $\Delta_{dBN}=80meV$. The calculation is the same as that in the dBN subsection but  $H_{0}$  is replaced with its renormalized version by $H_{V}$.  (c) The bands structure along cut1 as indicated in panel (b,d). (d) The Fermi surface corresponds to the band structure in (b). (e-g) illustrate the evolution of bands around M with Se height. }
\label{MDirac}
\end{figure}

Very recently, in addition to the splitting at $\Gamma$ and M, new ARPES experiments observed four Dirac cones around M in FeSe thin films \cite{DLFeng2015-1, ZXShen2015}. Previously, Dirac cones have been observed in BaFe$_2$As$_2$ and they were revealed to originate from band folding in SDW phase \cite{Richard2010}. Thus, the observed Dirac cones are particularly intrigue because of the absence of static magnetic order in FeSe. A recent DFT calculation has also predicted the existence of Dirac cones in FeSe but just in the predicted ``pair-checkerboard AFM'' (P-AFM) phase \cite{XGGong2015}. Due to band folding in P-AFM, the Dirac cone should be observed not only around M but also around $\Gamma$, which is inconsistent with experiment results. In the following, we will show that the observed Dirac cones can be explained by the dBN order with band renormalization from interatomic Coulomb interaction. First, we consider the renormalization from interatomic Coulomb interaction \cite{ZiQiangWang2015}. The renormalized bands are shown in Fig. \ref{MDirac}(a), where two Dirac cones labeled as $\Lambda$ are near $E_F$. Fig. \ref{MDirac}(b) shows the bands with dBN order in a twinned system, where $\Delta_{dBN}=80$ meV. We find that in one domain the Dirac cones on $\Gamma$-M$_x$ are pushed up and those on $\Gamma$-M$_y$ are pushed down and it is opposite in the second domain. The Dirac cone can be clearly seen in Fig. \ref{MDirac}(c), which is consistent with experiment. The corresponding Fermi surfaces are given in Fig. \ref{MDirac}(d), where four Dirac cones appear around M. They are consistent with those in experiment except the additional oval-shaped electron pockets at M. The ARPES results show that, as $T$ is lowered below $T_{nem}$,  the area of electron pockets at M is reduced but little change occurs on hole pockets. Since the number of electrons should be conserved in general phase transition, some electron pockets are not observed in the experiment. We argue that the missing pockets are just the two big elliptical electronic pockets at M.  The reason that they are not observed in ARPES is probably attributed to the photoemission matrix element effect.

With cobalt doping in multi-layer FeSe films, at first, the nematicity is suppressed significantly and then the Dirac cones disappear at a higher doping \cite{DLFeng2015-1}. The former effect attributes to the electron doping and the latter is the result of reduction of Se height induced by cobalt dopant. The dBN order is induced by the quantum fluctuations from the proximity of the Van Hove singularity ($E_{M1}$) to the Fermi level \cite{ZiQiangWang2015}. The electron doping moves the Fermi level away from Van Hove singularity, thus suppresses dBN order. This also explains the absence of nematic order in heavily electron doped 1UC FeSe on SrTiO$_3$ \cite{DLFeng2015-1}. Upon cobalt doping, Se height decreases and the $E_{M1}$ bands are pushed down and the $E_{M3}$ bands are pushed up. The critical case is that $E_{M1}$ bands meet $E_{M3}$ bands and the Dirac cones disappear, which is just the case of 8\% cobalt doping (see Fig. \ref{MDirac}(f)). If the doping further increases, $E_{M1}$ and $E_{M3}$ bands are inverted and an anticrossing between $\Sigma_{2'}$ and $\Sigma_{4}$ happens, resulting in a gap at M point. In this case, the band is similar to the band of 1UC FeSe \cite{JiangpingHu2014}. Fig. \ref{MDirac}(e-g) illustrates the bands around M with the decrease of Se height.

Based on the explanation of the observed Dirac cone and the effects of cobalt doping, two predictions can be made:  1) the dBN order in FeSe thin films may be enhanced with small hole doping. 2) the Dirac cones can survive and just exhibit an energy shift with dopants that can increase Se height or reduce Fe-Fe distance.

The driving force of nematicity in FeSe system is still under debate. Clear band splitting around M was discovered at the temperature that is much higher than the structure transition temperature. Furthermore, the formation of $C_2$ domain walls shows no correlation with lattice strain pattern \cite{ZXShen2015}. Therefore, the nematicity is unlikely related to lattice distortion.  Recent calculations show that interatomic Coulomb interaction can induce the dBN order \cite{ZiQiangWang2015}. Thus, the nematicity in FeSe system may have a electronic origin.

\textit{Summary: }
We investigated the effects on the electronic structure of eight possible orders in FeSe system. We found that only FO and C-AFM can simultaneously induce splittings at $\Gamma$ and M. B-AFM and SOC have very similar band structures on $\Gamma$-M near $E_{F}$. The $T$-insensitive splitting at $\Gamma$ and the $T$-dependent splitting at M can be explained by the dBN order together with SOC. The recent observed Dirac cones and their temperature dependence in FeSe thin films can also be well explained by the dBN order with band renormalization. Their thickness- and cobalt-doping- dependent behaviors are the consequences of electron doping and reduction of Se height. All these suggest that the nematic order in FeSe system is dominated by the dBN order, which is attributed to electronic origin.

{\it Acknowledgement: }
The work is supported by the Ministry of Science and Technology of China 973
program (Grant No. 2012CV821400 and No. 2010CB922904), National Science Foundation of China (Grant No. NSFC-1190024, 11175248 and 11104339), and   the Strategic Priority Research Program of  CAS (Grant No. XDB07000000).

\end{bibunit}

\begin{bibunit}

\clearpage
\pagebreak
\onecolumngrid
\widetext
\begin{center}
\textbf{\large Supplementary material for ``Effects of possible ordered phase on electronic structure of FeSe''}
\end{center}


\setcounter{equation}{0}
\setcounter{figure}{0}
\setcounter{table}{0}

 \renewcommand{\thefigure}{S\arabic{figure}}
 \renewcommand{\thetable}{S\arabic{table}}
 \renewcommand{\theequation}{S\arabic{equation}}

\section{Tight-binding model}
In our paper, the TB model $H_0$ is obtained from the five-orbital TB fit of  the DFT band structure for 1UC FeSe film with the lattice parameters of FeSe bulk, which is similar to the TB model for LaOFeAs given by Graser {\em et al.} except some additional hopping terms \cite{SGraser}.  The specific expressions are given  to facilitate others' use of it. $H_0$ can be written as
\begin{eqnarray}
H_{0}=\sum_{\mathbf{k}\in BZ1,\sigma}\psi_{\sigma}^{\dag}(\mathbf{k})A_0(\mathbf{k})\psi_{\sigma}(\mathbf{k}),
\end{eqnarray}
where $\psi^{\dag}_{\sigma}(\mathbf{k}) =[C_{\mathbf{k}1\sigma}^{\dag},C_{\mathbf{k}2\sigma}^{\dag},C_{\mathbf{k+Q}3\sigma}^{\dag},C_{\mathbf{k+Q}4\sigma}^{\dag},C_{\mathbf{k+Q}5\sigma}^{\dag}]$
, $\mathbf{Q}=(\pi,\pi)$ and the matrix elements of $A_0(\mathbf{k})$ are in the following:


\begin{eqnarray*}
e_{11/22}(\mathbf{k}) & = &
  \epsilon_{11/22}+2t_{x/y}^{11}cosk_{x}+2t_{y/x}^{11}cosk_{y}+4t_{xy}^{11}cosk_{x}cosk_{y}+2t_{xx/yy}^{11}cos2k_{x}+2t_{yy/xx}^{11}cos2k_{y}\\
 &  &
  +4t_{xyy/xxy}^{11}cosk_{x}cos2k_{y}+4t_{xxy/xyy}^{11}cos2k_{x}cosk_{y}+4t_{xxyy}^{11}cos2k_{x}cos2k_{y},\\
e_{33/44/55}(\mathbf{k}) & = &
  \epsilon_{33/44/55}+2t_{x}^{33/44/55}(cosk_{x}+cosk_{y})+4t_{xy}^{33/44/55}cosk_{x}cosk_{y}+2t_{xx}^{33/44/55}(cos2k_{x}+cos2k_{y})\\
 &  & +4t_{xxy}^{33/44/55}(cosk_{x}cos2k_{y}+cos2k_{x}cosk_{y})+4t_{xxyy}^{33/44/55}cos2k_{x}cos2k_{y},\\
e_{12}(\mathbf{k}) & = &
  -4t_{xy}^{12}sink_{x}sink_{y}-4t_{xxy}^{12}(sink_{x}sin2k_{y}-sin2k_{x}sink_{y})-4t_{xxyy}^{44}sin2k_{x}sin2k_{y},\\
e_{13/23}(\mathbf{k}) & = &
  \pm2it_{y}^{13}sink_{y/x}\pm4it_{xy}^{13}cosk_{x/y}sink_{y/x}\pm2it_{yy}^{13}sin2k_{y/x}\\
 &  & \pm4it_{xyy}^{13}cosk_{x/y}sin2k_{y/x}\pm4it_{xxy}^{13}cos2k_{x/y}sink_{y/x},\\
e_{14/24}(\mathbf{k}) & = &
  2it_{x}^{14}sink_{x/y}+4it_{xy}^{14}sink_{x/y}cosk_{y/x}+4it_{xxy}^{14}sin2k_{x/y}cosk_{y/x}\\
 &  & +2it_{xx}^{14}sin2k_{x/y}+4it_{xxyy}^{14}sin2k_{x/y}cos2k_{y/x},\\
e_{15/25}(\mathbf{k}) & = &
  2it_{x}^{15}sink_{y/x}-4it_{xy}^{15}cosk_{x/y}sink_{y/x}-4it_{xxyy}^{15}cos2k_{x/y}sin2k_{y/x}\\
 &  & +2it_{yy}^{15}sin2k_{y/x}+4it_{xyy}^{15}cosk_{x/y}sin2k_{y/x}+4it_{xxy}^{15}cos2k_{x/y}sink_{y/x},\\
e_{34}(\mathbf{k}) & = & 4t_{xyy}^{34}(sink_{x}sin2k_{y}-sin2k_{x}sink_{y}),\\
e_{35}(\mathbf{k}) & = &
   2t_{x}^{35}(cosk_{x}-cosk_{y})+4t_{xyy}^{35}(cosk_{x}cos2k_{y}-cos2k_{x}cosk_{y})+2t_{xx}^{35}(cos2k_{x}-cosk_{y}),\\
e_{45}(\mathbf{k}) & = & 4t_{xy}^{45}sink_{x}sink_{y}+4t_{xxyy}^{45}sin2k_{x}sink_{y}.
\end{eqnarray*}

\begin{table}[h]
  \centering
  \caption{The on-site energy used for the DFT fit of the five-orbital TB model.}
  \begin{tabular}{ccccc}
   \hline
   $\epsilon_{xz}$ & $\epsilon_{yz}$ & $\epsilon_{x^{2}-y^{2}}$ & $\epsilon_{xy}$ & $\epsilon_{z^{2}}$\tabularnewline
   \hline
   0.050 & 0.050 & -0.483 & -0.035 & -0.403\tabularnewline
   \hline
  \end{tabular}
\end{table}

\begin{table}[h]
  \centering
  \caption{The intraorbital hopping parameters used for the DFT fit of the five-orbital TB model.}
  \begin{tabular}{ccccccccc}
\hline
$t_{i}^{mm}$ & i=x & y & xy & xx & yy & xxy & xyy & xxyy\tabularnewline
\hline
m=1 & -0.069 & -0.317 & 0.227 & 0.002 & \emph{-0.036} & -0.019 & 0.014 & 0.024\tabularnewline
m=3 & 0.396 &  & -0.070 & -0.013 &  &  &  & 0.012\tabularnewline
m=4 & 0.061 &  & 0.085 & 0.002 &  & -0.019 &  & -0.024\tabularnewline
m=5 & 0.005 &  & 0.013 & -0.014 &  & 0.006 &  & -0.011\tabularnewline
\hline
\end{tabular}
\end{table}

\begin{table}[h]
  \centering
  \caption{The intraorbital hopping parameters used for the DFT fit of the five-orbital TB model.}
  \begin{tabular}{cccccccc}
\hline
$t_{i}^{mn}$ & i=x & xy & \emph{xx} & \emph{yy} & xxy & \emph{xyy} & xxyy\tabularnewline
\hline
mn=12 &  & 0.103 &  &  & -0.011 &  & 0.032\tabularnewline
mn=13 & 0.380 & -0.089 & -0.011 &  & -0.018 & 0.006 & \tabularnewline
mn=14 & 0.306 & 0.053 & -0.001 &  & 0.006 &  & \emph{-0.009}\tabularnewline
mn=15 & 0.158 & 0.130 &  & 0.009 & -0.009 & -0.011 & 0.012\tabularnewline
mn=34 &  &  &  &  & -0.012 &  & \tabularnewline
mn=35 & -0.329 &  & -0.023 &  & -0.006 &  & \tabularnewline
mn=45 &  & 0.113 &  &  &  &  & -0.011\tabularnewline
\hline
\end{tabular}
\end{table}

\section{The Calculation of the $\Gamma$ splitting in C-AFM order}
 C-AFM induces an additional Hamiltonian term,
\begin{equation}
h^{CA} = \Delta_{CA}\sum_{\mathbf{k}\in BZ1,\alpha,\sigma}
  sgn(\sigma)C_{\mathbf{k},\alpha,\sigma}^{+}C_{\mathbf{k}+\mathbf{q}_{CA},\alpha,\sigma}.
\end{equation}
where $sgn(\sigma)$ is the sign function, equals 1 for spin up and -1 for spin down.
Define $\varphi_{CA,\sigma}(\mathbf{k})^{+}=[\psi_{\sigma}^{+}(\mathbf{k}),\psi_{\sigma}^{+}(\mathbf{k}+\mathbf{q}_{CA})]$,
then the total Hamiltonian
$H_{0}^{CA}=\sum_{\mathbf{k}\in\{BZ_{1Fe},kx\geqslant0\}}
   \varphi_{CA,\sigma}(\mathbf{k})^{+}A_{\sigma}^{CA}(\mathbf{k})\varphi_{CA,\sigma}(\mathbf{k})$,
where
\[
A_{\sigma}^{CA}(\mathbf{k})=\left[\begin{array}{cc}
A^{nor}(\mathbf{k}) & sgn(\sigma)\Delta_{CA}\mathbf{I}\\
sgn(\sigma)\Delta_{CA}\mathbf{I} & A^{nor}(\mathbf{k}+\mathbf{q}_{CA})
\end{array}\right],
\]
and $\mathbf{I}$ is an $5\times5$ identity matrix.

The splitting of $E_{g}$ state can be exactly solved,

\begin{equation}
\Delta_{E_{g}}^{CA} =  e_{22}(\pi,0)-e_{11}(\pi,0)+\sqrt{(E_{g}-e_{22}(\pi,0))^{2}+4\Delta_{CA}^{2}}
     -\sqrt{(E_{g}-e_{11}(\pi,0))^{2}+4\Delta_{CA}^{2}}\propto\Delta_{CA}^2.
\end{equation}

\section{The Calculation of the $\Gamma$ splitting in B-AFM order}

In B-AFM phase, the additional Hamiltonian terms is given by
\begin{equation}
h^{BA} = \Delta_{BA}\sum_{k\in BZ_{1Fe}}
(\frac{1-i}{2}C_{\mathbf{k}\alpha\sigma}^{+}C_{\mathbf{k}+\mathbf{q}_{BA}\alpha\sigma}
   +\frac{1+i}{2}C_{\mathbf{k}\alpha\sigma}^{+}C_{\mathbf{k}-\mathbf{q}_{BA}\alpha\sigma})sgn(\sigma).
\end{equation}
Define $\varphi_{BA,\sigma}(\mathbf{k})^{+}
   =[\psi_{\sigma}^{+}(\mathbf{k}),\psi_{\sigma}^{+}(\mathbf{k}+\mathbf{Q}),
     \psi_{\sigma}^{+}(\mathbf{k}-\mathbf{q}_{BA}),\psi_{\sigma}^{+}(\mathbf{k}+\mathbf{q}_{BA})]$,
then the total Hamiltonian reads $H^{BA}=\sum_{\mathbf{k}\in BZ_{BA}}\varphi_{BA,\sigma}(\mathbf{k})^{+}A_{\sigma}^{BA}(\mathbf{k})\varphi_{BA,\sigma}(\mathbf{k})$,
where $BZ_{BA}$ is the magnetic BZ for B-AFM phase and
\[
A_{\sigma}^{BA}(\mathbf{k})=
\left[\begin{array}{cccc}
A_{0}(\mathbf{k}) & 0 & D_{-} & D_{+}\\
0 & A_{0}(\mathbf{k}+\mathbf{Q}) & D_{+} & D_{-}\\
D_{-}^{+} & D_{+}^{+} & A_{0}(\mathbf{k-q_{BA}}) & 0\\
D_{+}^{+} & D_{-}^{+} & 0 & A_{0}(\mathbf{k+q_{BA}})
\end{array}\right],
\]
here $D_{\pm}=\frac{1\mp i}{2}\Delta_{BA}sgn(\sigma)\mathbf{I}$.

B-AFM leads to the splitting at $\Gamma$, which can be evaluated with Perturbation theory.  $A_{\sigma}^{BA}(\Gamma)-A_{\sigma}^{BA}(\Gamma,\Delta_{BA}=0)$ is taken as a perturbation, denoted as $V^{BA}$. With regular first- and second- order perturbation formulas, no splitting is found. The indirect
hopping between $C_{\mathbf{\Gamma},xz,\sigma}^{+}$ and $C_{\Gamma,yz,\sigma}$ through $C_{\mathbf{\pm q}_{BA},\alpha,\sigma}$ need including in the first order perturbation matrix \cite{LandauQM}. The elements of the perturbation matrix for $E_{g}$ states are

\begin{equation}
V_{i,j}^{BA} = \frac{1}{2}\Delta_{BA}^{2} \sum_{m=1}^{5}
             \frac{\phi_{m,-\mathbf{q_{BA}}}(i)\phi_{m,-\mathbf{q_{BA}}}(j)^{*}+\phi_{m,\mathbf{q_{BA}}}(i)\phi_{m,\mathbf{q_{BA}}}(j)^{*}}
             {E_{g}-E_{m}(\mathbf{q}_{BA})}
\end{equation}
where $i,j=1,2$, $\phi_{m,\mathbf{\pm q_{BA}}}(i)$ is the $i$-th component of $m$-th eigenvector of $A_{0}(\mathbf{q}_{BA})$ and its corresponding eigenvalue is $E_{m}(\mathbf{q}_{BA})$. The splitting $\Delta_{E_{g}}$ can be calculated out by diagonalizing $\{V_{i,j}^{BA}\}_{2\times2}$. $\Delta_{E_{g}}=0.36\Delta_{BA}^{2}$ in our model.

\section{The Calculation of the $\Gamma$ splitting induced by SOC}
we consider an isotropic on-site SOC, $h^{SOC}=\lambda\sum_{i}\mathbf{L}_{i}\mathbf{S}_{i}$ where $\sum_{i}$ sums over the Fe sites.
In $3d-$orbital space, $h^{SOC}$ is written as

\begin{eqnarray}
h^{SOC} & = & \frac{1}{2}\lambda\sum_{k\in BZ1,\sigma}\Bigl\{ i (C_{k,xz,\sigma}^{+}C_{k,xy,\bar{\sigma}}
 -C_{k,yz,\sigma}^{+}C_{k,x^{2}-y^{2},\bar{\sigma}}-\sqrt{3}C_{k,yz,\sigma}^{+}C_{k,z^{2},\bar{\sigma}})\nonumber\\
  &  & -sgn(\sigma)(C_{k,yz,\sigma}^{+}C_{k,xy,\bar{\sigma}}+C_{k,xz,\sigma}^{+}C_{k,x^{2}-y^{2},\bar{\sigma}}
  -\sqrt{3}C_{k,xz,\sigma}^{+}C_{k,z^{2},\bar{\sigma}})\nonumber\\
 &  & -i sgn(\sigma)(C_{k,xz,\sigma}^{+}d_{k,yz,\sigma}+2C_{k,x^{2}-y^{2},\sigma}^{+}C_{k,xy,\sigma})+H.c.\Bigr\}.
\end{eqnarray}

Define $\varphi(\mathbf{k})^{+}=[\psi_{\uparrow}^{+}(\mathbf{k}),\psi_{\downarrow}^{+}(\mathbf{k}+\mathbf{Q})],$
then the total Hamiltonian $H_{0}^{SOC}=H_{0}+h^{SOC}=\sum_{\mathbf{k}\in BZ1}\varphi(\mathbf{k})^{+}A^{SOC}(\mathbf{k})\varphi(\mathbf{k})$,
where
\begin{eqnarray}
A^{SOC}(\mathbf{k})& = & \left[\begin{array}{cc}
    A_{0}(\mathbf{k})+h_{\uparrow\uparrow}^{soc} & h_{\uparrow\downarrow}^{soc}\\
    h_{\uparrow\downarrow}^{soc+} & A_{0}(\mathbf{k}+\mathbf{Q})-h_{\uparrow\uparrow}^{soc}
  \end{array}\right]
\\
h_{\uparrow\uparrow}^{soc} & =& \frac{\lambda}{2}\left[\begin{array}{ccccc}
 & -i&\\
i\\
 &  &  & -2i\\
 &  & 2i\\
\\
  \end{array}\right]
\\
h_{\uparrow\downarrow}^{soc} & = & \frac{\lambda}{2}\left[\begin{array}{ccccc}
 &  & -1 & i & \sqrt{3}\\
 &  & -i & -1 & -\sqrt{3}i\\
 1 & i\\
-i & 1\\
-\sqrt{3} & \sqrt{3}i
\end{array}\right].
\end{eqnarray}
The $E_{g}$ level at $\Gamma$,is splitted into two energies,$E_{g\pm}$ by SOC. $A^{SOC}(\mathbf{k})-A^{SOC}(\mathbf{k},\lambda=0)$ is taken as the perturbation and then with the regular perturbation theory and accurate to the second order terms of $\lambda$,

\begin{eqnarray}
E_{g+} & = & E_{g}+\frac{1}{2}\lambda+\frac{\lambda^{2}}{2(E_{g}-e_{33}(\Gamma))}
         +\frac{\lambda^{2}}{2(E_{g}-e_{44}(\Gamma))},\\
E_{g-} & = & E_{g}-\frac{1}{2}\lambda+\frac{3\lambda^{2}}{2(E_{g}-e_{55}(\Gamma))},
\end{eqnarray}
where $e_{\alpha\beta}(\mathbf{k})$ is the element of $A_{0}(\mathbf{k})$. Thus, the splitting  $\Delta_{E_g}$ is,

\begin{eqnarray}
 \Delta_{E_g} & = & \lambda+\frac{\lambda^{2}}{2(E_{g}-e_{33}(\Gamma))}
         +\frac{\lambda^{2}}{2(E_{g}-e_{44}(\Gamma))}-\frac{3\lambda^{2}}{2(E_{g}-e_{55}(\Gamma))}.
\end{eqnarray}

\end{bibunit}

\end{document}